\documentclass[twocolumn,preprintnumbers,amsmath,amssymb,superscriptaddress]{revtex4}
\usepackage{amsmath,amsfonts,amssymb}
\usepackage{color}

\usepackage{amsmath,amsfonts,amssymb}
\usepackage[english]{babel}
\usepackage[T1]{fontenc}
\usepackage{color}
\usepackage{float}
\usepackage{verbatim}
\usepackage{graphicx}
\usepackage{bm}
\usepackage{mathtools}
\usepackage{stmaryrd}
\usepackage{anyfontsize}
\usepackage{changepage}

\usepackage{natbib}
\definecolor{darkblue}{rgb}{0,0,0.6}
\definecolor{darkcyan}{rgb}{0.1,0.3,0.4}
\definecolor{darkgreen}{rgb}{0,0.4,0}
\definecolor{darkred}{rgb}{0.6,0,0}

\newcommand{\beginsupplement}{%
        \clearpage
        \setcounter{table}{0}
        \renewcommand{\thetable}{S\arabic{table}}%
        \setcounter{figure}{0}
        \renewcommand{\thefigure}{S\arabic{figure}}%
     }

\begin{document}

\title{Scaling of the risk landscape drives optimal life history strategies and the evolution of grazing} 

\author{Uttam Bhat} \affiliation{School of Natural Sciences, University
  of California, Merced, Merced, CA 95340, USA}

\author{Christopher P. Kempes} \affiliation{Santa Fe Institute, 1399 Hyde Park Road, Santa Fe, NM 87501, USA}

\author{Justin D. Yeakel} \affiliation{School of Natural Sciences, University
  of California, Merced, Merced, CA 95340, USA}


\begin{abstract}
  Consumers face numerous risks that can be minimized by incorporating different life-history strategies.
  How much and when a consumer adds to its energetic reserves or invests in reproduction are key behavioral and physiological adaptations that structure much of how organisms interact.
  Here we develop a theoretical framework that explicitly accounts for stochastic fluctuations of an individual consumer's energetic reserves while foraging and reproducing on a landscape with resources that range from uniformly distributed to highly clustered.
  First, we show that optimal life-history strategies vary in response to changes in the mean productivity of the resource landscape, where depleted environments promote reproduction at lower energetic states, greater investment in each reproduction event, and smaller litter sizes.
  We then show that if resource variance scales with body size due to landscape clustering, consumers that forage for clustered foods are susceptible to strong Allee effects, increasing extinction risk.
  Finally, we show that the proposed relationship between consumer body size, resource clustering, and Allee effect-induced population instability offers key ecological insights into the evolution of large-bodied grazing herbivores from small-bodied browsing ancestors.\\ \\
  Life-history strategies $|$ Foraging dynamics $|$ Population dynamics $|$ Evolution of Grazing
\end{abstract}

\maketitle


\vspace{3mm}
\begin{adjustwidth}{2.5em}{0em}
\textbf{Significance} Consumer species assume diverse life history and foraging strategies in part to mitigate the risks imparted by spatially variable resources.
By deriving a mechanistic model of energy allocation, we show how fitness optimizing strategies are tied to resource variability, and that population stability depends on the scaling of resource variability with consumer body size and diet. 
These relationships offer insight into the evolutionary trend towards larger body size known as Cope's rule, and the mammalian transition from browsing to grazing following the advent of grasslands in the mid-late Miocene.\\
\end{adjustwidth}



The landscape of risk faced by consumers is determined not only by the mean density of potential foods but their variability over both space and time \cite{Mangel1986}. 
Consumer behavioral and life history strategies are selected for or against over evolutionary time in part to manage these risks \cite{Roff2002}.
At a coarse scale, these strategies involve how and when energy is saved either endogenously or exogenously \cite{Gerber2004}, and when it is spent.
For many species, the most substantial metabolic rate expenditures are those incurred during reproduction \cite{Yeakel_Dynamics_of_Starvation_and_Recovery}, and this is particularly true for endotherms, which on average spend more energy per offspring than non-endothermic organisms \cite{Lindstedt1981}.
Both the availability and variability of resources interact directly with the physiological and metabolic constraints of the consumer to give rise to the remarkable diversity of foraging and life history strategies observed in nature \cite{Burke_Central_place_foraging_response_to_fluctuations,Schmickl_Cost_of_Fluctuations_in_Honey_Bees,Brown_Marquet_Evolution_of_Body_Size,Grunbaum2012,Yeakel_Dynamics_of_Starvation_and_Recovery,Charnov_Optimal_foraging,MacArthur_Pianka_On_Optimal_Use_of_a_Patchy_Environment}.

A consumer must enact a fitness-maximizing strategy that differentially allocates energy to somatic growth, maintenance, and reproduction under the constraints imposed by the spatial distribution of available resources \cite{Yeakel_Dynamics_of_Starvation_and_Recovery}. 
How the energy gathered from resources must be invested into alternative pathways to serve different functions is often referred to as the Y tradeoff \cite{vanNoordwijk1986}.
Theoretical investigations of life history strategies are generally explored using optimization models, where different combinations of energetic or risk trade-offs are used to examine the influence of a trait, or set of traits, on some measure of fitness \cite{Stearns1989,Roff2002}. 
Moreover, optimal life history strategies are often estimated assuming deterministic metabolic and growth rates \cite{Charnov_1991_Life_history_variation}, however growth rates are known to be flexible depending on the availability of resources and the age-structure of mortality \cite{Abrams2001}.
Because the variability in resource acquisition alters both individual growth rates and mortality due to starvation, it should -- by extension -- influence the selection of alternative life histories.
Resource variability scales differently with consumer size and diet, meaning a single landscape can impart a wide diversity of experiences and challenges to different organisms \cite{Grunbaum2012}.
For instance, a savanna is a vast forest to the common African rat (\emph{Mastomys natalensis}), where grass seeds that comprise a large portion of its diet are unevenly dispersed in discrete units across a single patch that might represent a sizable fraction of its home-range \cite{Jarvis2003}.
To larger herbivores such as impala (\emph{Aepyceros melampus}), grasses are resources that are more or less homogeneously distributed across discrete patches \cite{Fryxell2005}, and to megafauna such as wildebeest (\emph{Connochaetes} spp.), such foods appear even more uniform over space.
Within the same landscape, the clustering of fruits such as wild cucumber or figs may not change drastically with increasing frugivore body size (cf. Ref. \citenum{Malmborg1988}), meaning that the foraging challenges facing these consumers may be similar, regardless of size. 

As with frugivores, carnivores also specialize on resources that are concentrated in nutritionally rich but spatially distributed units, however because prey size scales strongly with consumer size, so does its density across the landscape \cite{Hatton2015}. 
For example, small-bodied carnivores such as serval (\emph{Leptailurus serval}) face a relatively uniform resource landscape (primarily rodents) compared to leopards (\emph{Panthera pardus}) that face a landscape where resources (larger herbivores) are more clustered. 
Overall, the greatest amount of resource clustering could be experienced by either the smallest or largest consumers, depending on the resource type and the variability in distances between resources across the landscape. 



The variability of risks and rewards has large effects on the expected optimality of life history strategies \cite{Roff2007}, as well as implications for population dynamics \cite{Moore2014}. 
While the uncertainty associated with foraging payoffs is clearly important, it is the mean effect in lieu of variability that is generally considered when exploring the constraints giving rise to alternative life histories. 
Mean-field models of resource consumption based on average interaction rates \cite{jetz2004scaling,DeLong2012a,DeLong2012b,Yeakel_Dynamics_of_Starvation_and_Recovery}, and the explicit incorporation of resource distributions across fractal landscapes \cite{haskell2002fractal}, have been used to explore the population-level effects of scaling home range size and/or population density with body size. 
Even the dimensionality of resource landscapes can have a significant effect on consumer acquisition and consumption rates, and important consequences for population dynamics \cite{Pawar_2012_dimensionality}.
In contrast, the influence of resource variance in determining optimal life histories, and the population-level consequences of these strategies, has received relatively little attention.

Here we explore the effects of both resource availability and variability on the optimality of life history strategies with a mechanistic model of energy allocation.
Using stochastic process theory \cite{Redner2014}, we model the dynamics of an individual consumer's energetic state over time as a function of mean energetic gains, the variance of those gains, metabolic losses, and reproductive investment.
We examine the energetic dynamics of an individual consumer based on well-established constraints of foraging and reproduction among terrestrial mammals, provide predictions on a suite of life history characteristics, and then extend our methodology to explore population-level implications.


Our results offer four fundamental insights into the effects of resource availability and variability in driving the evolution of optimal life history strategies.
First, we show that optimal strategies vary in response to changes in the mean productivity of the resource landscape, where depleted environments promote reproduction at lower energetic states, a greater investment in each reproduction event, and smaller litter sizes.
Second, we show that including resource variability allows for predictions of life-history strategies with optimal population sizes deviating from the negative $3/4$-power scaling of Damuth's law \cite{Damuth1987}.
Third, we show that by integrating different scalings of resource variability with consumer body size, population densities of species that forage for more clustered foods, such as frugivores or carnivores, are more sensitive to increasing resource variability than those that consume more homogenously distributed foods such as grazers.
A key result of our analysis reveals that consumers specializing on resources that are clustered are prone to strong Allee effects, which increases extinction risks.
Finally, we show that the proposed relationship between consumer body size, resource clustering, and the population instabilities arising from these Allee effects, provides an ecological mechanism for the evolution of large-bodied grazing herbivores from small-bodied browsing ancestors, a well-documented transition in mammalian evolution following the advent of grasslands in the mid-late Miocene.


\section*{Predicting optimal life history strategies}
\label{sec:freeopt}
\begin{figure}
\centering
\includegraphics[width=0.4\textwidth]{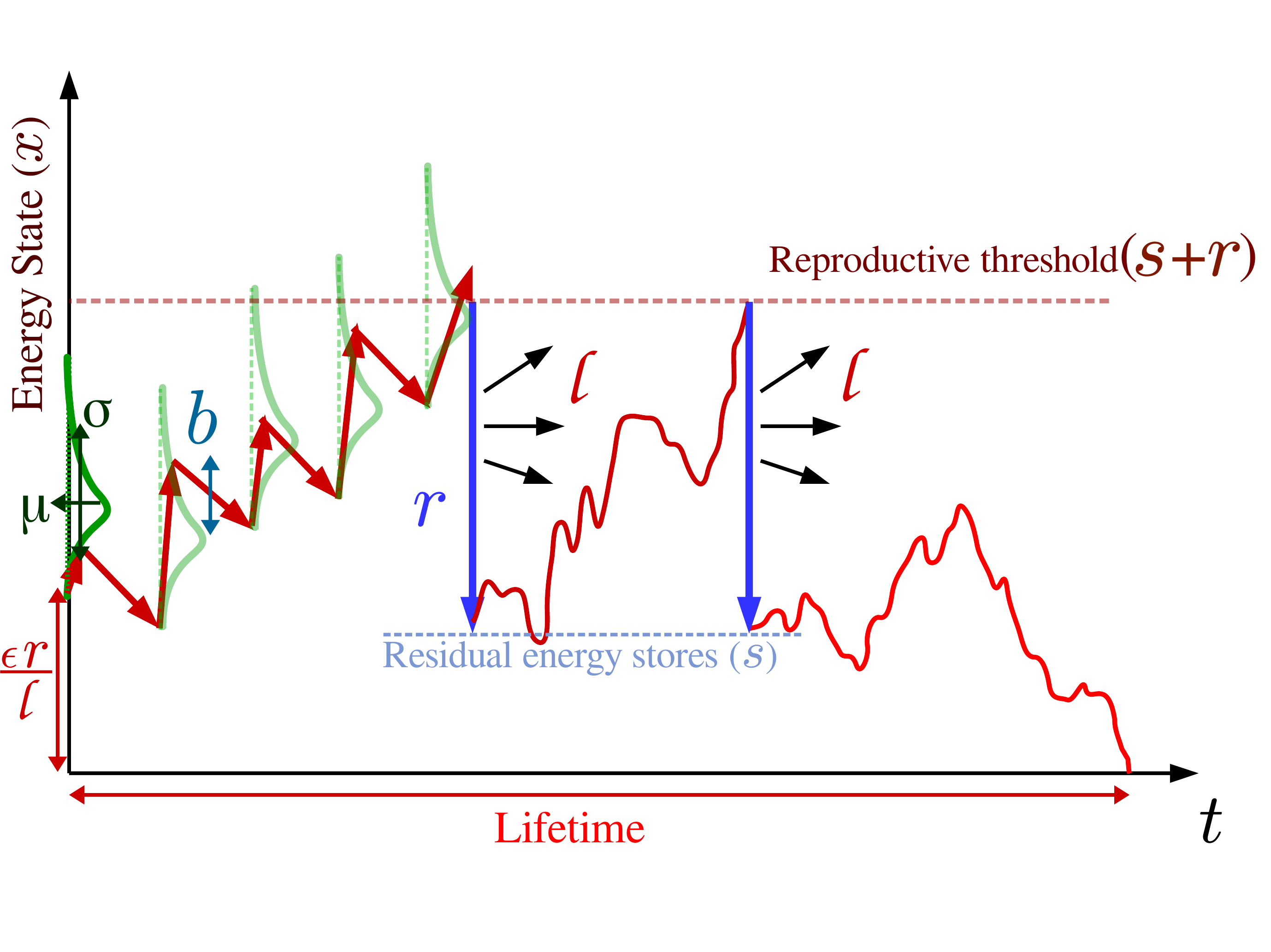}
\caption{A schematic of the individual energetic dynamic framework and 1) energetic investment in each reproductive event, $r$, 2) energetic investment in somatic reserves, $s$, and 3) litter size, $\ell$.}
\label{model-illustration}
\end{figure}

We model the dynamics of an individual consumer's energetic state over time as a function of energetic gains from acquiring resources, and costs from metabolic losses and reproductive investment.
For simplicity, the consumer's energetic state is described by a single state variable \cite{Mangel_Unified_Foraging_Theory}, $X=x$, which measures the onboard energetic stores available for both metabolic (ontogenetic growth and maintenance) and reproductive expenditures (Fig. \ref{model-illustration}).
Here and henceforth, we use uppercase notation for stochastic variables and lowercase notation for specific values of stochastic quantities.
The consumer's energetic state increases by the amount of food it obtains in a day $G=g$, which is stochastic and normally distributed with mean $\mu$ and variance $\sigma^2$. 
The consumer's energetic state decreases by a fixed daily metabolic cost associated with somatic maintenance $b$, regardless of foraging success.
The organism dies of starvation when the state drops to zero.
Energetic investment in reproduction occurs at the threshold $X = s+r$, where $r$ is the energetic cost of reproduction and $s$ is the minimal somatic reserves that must be maintained during gestation.
The consumer's post-reproductive energetic state is reduced to $X=s$, and the amount spent on reproduction is equally partitioned among $\ell$ offspring within the litter, with efficiency $\epsilon$.
The specific values of the variables $r$, $s$, and $\ell$ thus define the life history strategy of the organism.

We measure the fitness of the organism by the total number of surviving offspring produced during its lifetime, $W$, which is the generational replacement rate and can be converted to the specific growth rate using the known time to reach reproductive age \cite{Yeakel_Dynamics_of_Starvation_and_Recovery}.
The values of $r$, $s$, and $\ell$ impact fitness by adjusting the amount of investment in and timing of, each reproductive event, which in turn impacts the number and success of reproductive events in a lifetime (Fig.~\ref{model-illustration}).
By accounting for differential energetic allocation in reproductive versus somatic investments, we mechanistically incorporate the trade-offs thought to play a central role in determining life history strategies.
We note that our model does not take into account external mortality such as predation, nor an intrinsic death rate, however these rates are expected to be small compared with starvation-induced mortality \cite{Yeakel_Dynamics_of_Starvation_and_Recovery}.
 
\begin{figure*}[t]
\centering
\includegraphics[width=0.4\textwidth]{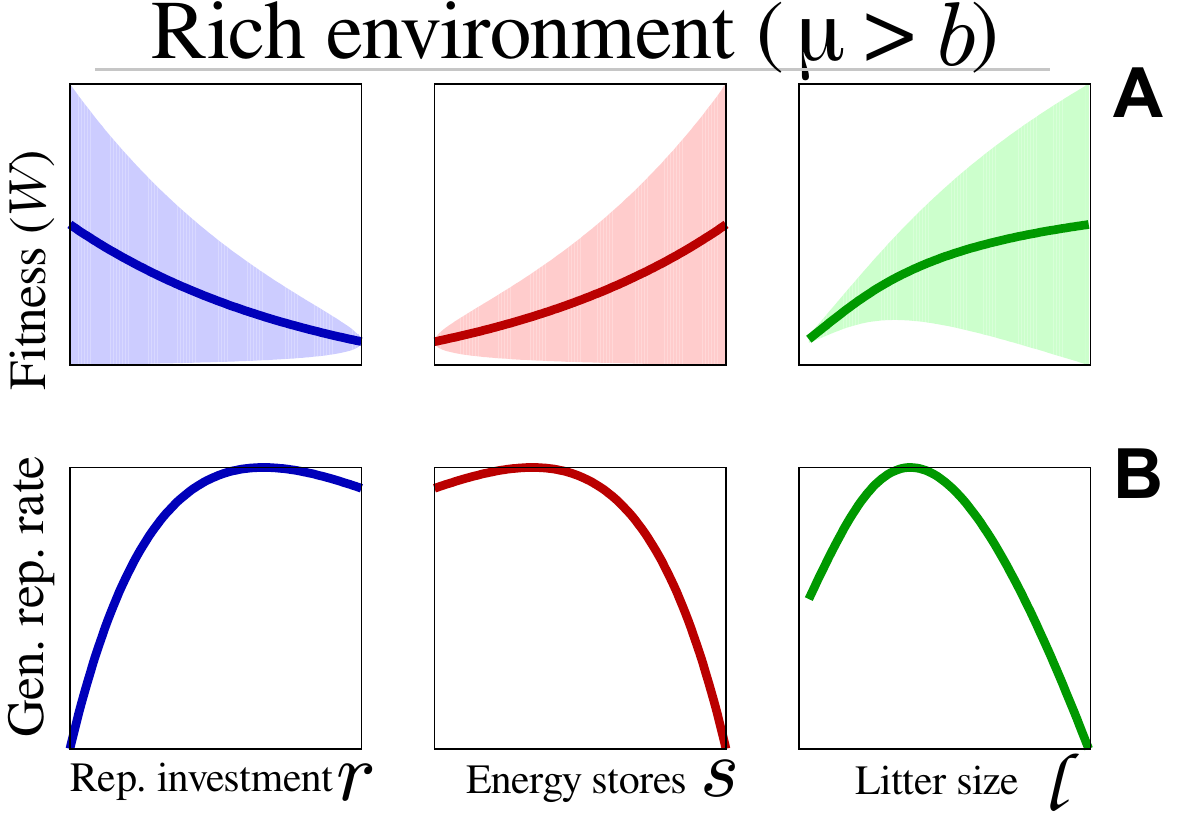}
\includegraphics[width=0.4\textwidth]{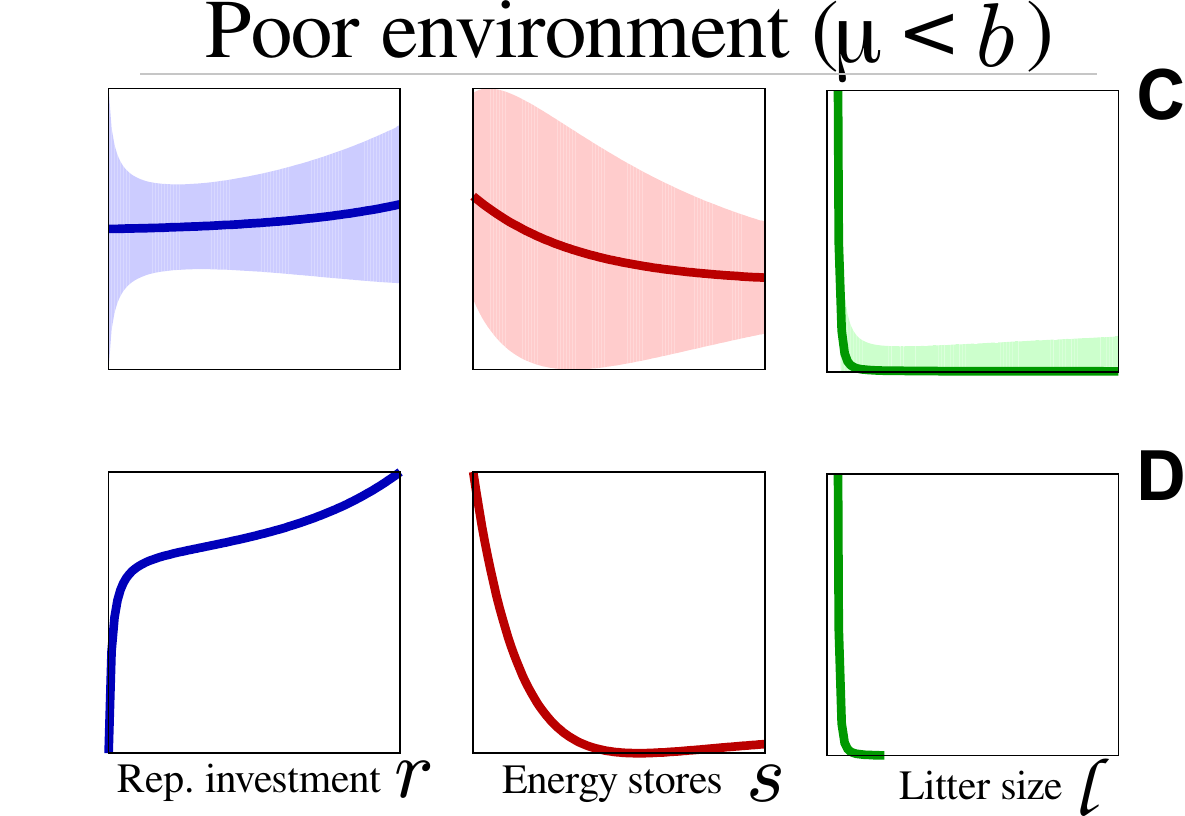}
\caption{Qualitative plots of reproductive fitness with the various parameters of the optimization problem.}
\label{freeopt-results}
\end{figure*}

We first treat our framework as an unconstrained optimization problem, where the goal is to maximize the number of surviving offspring across life history variables $r$, $s$, and $\ell$.
Fitness maximization of life history variables is carried out with respect to four parameters:
\emph{i}) the average amount of food obtained per day $\mu$,
\emph{ii}) the variance of food obtained per day $\sigma^2$,
\emph{iii}) the daily energetic cost of metabolism $b$, and
\emph{iv}) the energetic transfer efficiency from parent to offspring $\epsilon$.
The dependence of average fitness, $\left\langle W \right\rangle$ on the parameters and variables described above, can be solved analytically (see SI Appendix A) to give the expression
\begin{equation}
 \left\langle W \right\rangle = \frac{\ell\left(1-{\rm e}^{-2(\mu-b)\epsilon r/\ell \sigma^2}\right)}{{\rm e}^{-2(\mu-b)s/\sigma^2}-{\rm e}^{-2(\mu-b)(s+r)/\sigma^2}}\,.
\end{equation}
The dependence of $\left\langle W \right\rangle$ for the unconstrained model is shown in Fig.~\ref{freeopt-results}A \& \ref{freeopt-results}C.
Because the number of offspring is stochastic, the population growth rate can be estimated by $\left\langle W \right\rangle - \mathrm{Var}(W)/2$ (Fig.~\ref{freeopt-results}B \& \ref{freeopt-results}D), which is a first-order approximation of the geometric mean \cite{Hilborn_1997_Ecological_Detective} (see SI Appendix A for the expression for $\mathrm{Var}(W)$).
The population growth rate thus depends on both the fitness mean and variance, both of which depend on the availability and variability of resources to the consumer.
We observe that fitness variance plays an important role in the qualitative results of our model: although mean fitness may be high with respect to specific values of life history variables $r$, $s$, or $\ell$, an associated high variance in fitness serves to penalize the expected population growth rate.
This effect results in fitness optima at intermediate values of life history variables that describe somatic and reproductive investment strategies.

Calculation of the population growth rate reveals fitness peaks that depend strongly on resource availability.
When the average rate of obtaining food is greater than daily metabolic costs (i.e., $\mu>b$), we observe that the fitness-maximizing strategy is to 
\emph{i}) decrease investment in each reproductive event (low $r$),
\emph{ii}) increase the minimal somatic reserves maintained post-reproduction (high $s$), and
\emph{iii}) increase the litter size for each reproductive event (high $\ell$) (Fig.~\ref{freeopt-results}A \& \ref{freeopt-results}B.).
In words, when environmental conditions are very rich, our model predicts that the optimal life history approach skews towards an `r-selection' strategy, where very little investment is partitioned among a large number of offspring (c.f. \cite{Reznick2002}).
Moreover, because these conditions also result in higher post-reproduction somatic reserves $s$, iteroparity is expected to maximize fitness.

When the average rate of obtaining food is less than daily metabolic costs (i.e., $\mu<b$), the environment is extremely poor, such that every day would be expected to result in energetic losses.
Although extreme, this condition illustrates how the qualitative predictions of the model depend on access to resources.
Because every reproductive opportunity may be the last in environments where resources are scarce, our model predicts increased reproductive investment (high $r$) in a minimal number of offspring (low $\ell$) (Fig.~\ref{freeopt-results}C \& \ref{freeopt-results}D.).
Poor environments also result in a minimal post-reproductive somatic reserves $s$, meaning that all of an organism's resources are invested in a single reproductive event (semelparity).
Taken together these findings agree with previous observations that increased investment in each reproductive event (higher $r$) is correlated with less investment in future reproductive events (lower $s$) \cite{Crespi2002}.

Our predictions overlap partially but not completely with classic expectations of optimal life history strategies in rich and poor environments \cite{Pianka1970,Reznick2002}.
Resource-limitation is expected to result in greater investment in fewer offspring, which corresponds to our model predictions of higher $r$ and lower $\ell$.
However we also predict that poor environments may promote semelparity, which goes against prior expectations.
In our framework, which exclusively considers resource availability and variability, semelparity arises when the organism invests maximally in a single reproduction at the expense of post-reproductive somatic reserves (low $s$), increasing the likelihood of subsequent starvation-induced mortality.
Thus, in extremely resource-limited environments, all effort is directed towards the first reproductive event because the likelihood of a second is so low.

Our unconstrained model also predicts fractional litter sizes $\ell < 1$ in resource-limited environments, meaning that the optimal strategy is to invest in less than one offspring \emph{per-capita}.
In such cases, multiple adults are required to invest in a single offspring, a process mirroring inclusive fitness among social organisms \cite{Rood_1990_Group_size_survival_reproduction}.
In fact, there is some evidence to suggest that resource limitation may promote the evolution of sociality \cite{Cahan2002}.
While our framework is minimal, it is tempting to speculate that the dynamics explored here may contribute to the evolutionary pressures selecting for such cooperative behaviors. 

So far, we have ignored the feedback that must exist between individual reproductive fitness and the availability of resources available to each consumer. 
As a high rate of reproduction will lead to an increasing population feeding on a finite resource, individual fitness is expected to decline. 
To understand this feedback and, consequently, the evolutionary endpoints of alternative life-history strategies, we need to ascertain the population-level fitness of the consumer. 
We next expand our framework to incorporate the effects of a self-limiting consumer population feeding on finite resources and examine to what extent resource availability and variability impacts the size and stability of consumer populations.

\section*{Effects on population stability}
A population is composed of multiple individuals that must partition available resources.
Because resource availability in part determines fitness, individual fitness is thus a function of population density.
We now examine how differences in the availability and variability of resources impacts the steady-state population densities as a function of different life history strategies determined by $r$, $s$, and $\ell$.
We first assume that the mean and variance of resources, $\mu$ and $\sigma$ respectively, are equally partitioned among $n$ members of a population as
\begin{equation}
\mu_{\text{ind}} = \mu/n\,,\qquad \sigma^2_{\text{ind}} = \sigma^2/n^\zeta,
\end{equation}
where resource variance follows a Taylor's power law relationship \cite{Taylor1961}.
As such, $\zeta$ determines how the resource variance experienced by an individual consumer scales with the size of the population, and this will depend on both the consumer's resource distribution as well as the area over which it forages. 


\begin{figure}[t]
\centering
\includegraphics[width=0.45\textwidth]{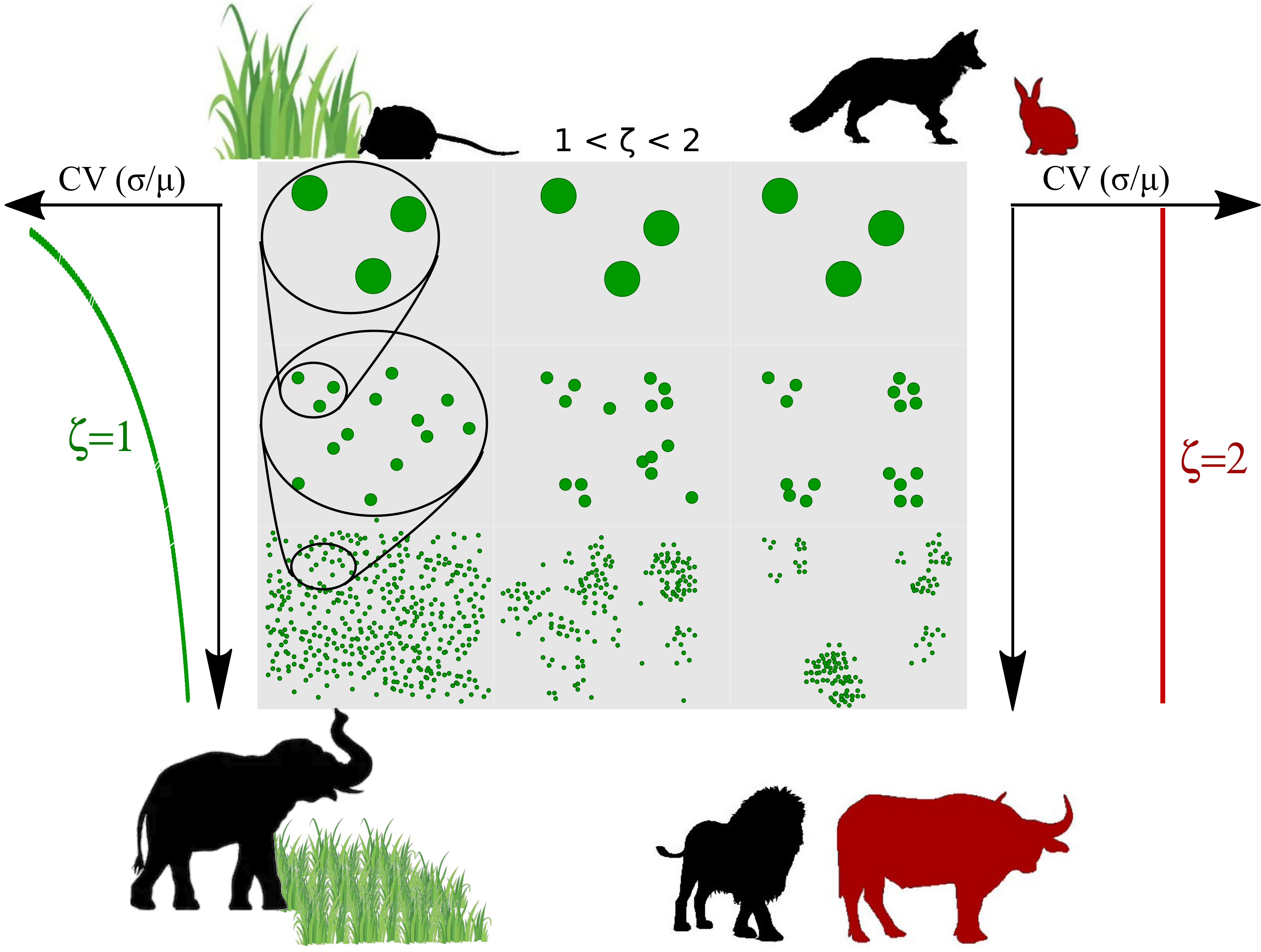}
\caption{A visual representation of resources with different variance-scaling exponent, $\zeta$. At $\zeta=1$, resource distribution is homogeneous (left-most panels), i.e., larger consumers face less spatial stochasticity (smaller CV) compared to smaller consumers. Whereas at $\zeta=2$, resource distribution is highly clustered (right-most panels). This could be due to spatial distribution of food, or due to size of individual units being proportionally larger for a larger consumer}
\label{zeta-illustration}
\end{figure}

While resources are partitioned equally, the variance of subdivided resources depends on its spatial distribution.
At one extreme, resources that are more uniformly distributed (e.g. grasses), are represented by $\zeta = 1$ (left-most panel of Fig.~\ref{zeta-illustration}).
For $\zeta>1$, there exists a spatial correlation in the resource distribution with explicit clustering at different spatial scales (center and right panels of fig.~\ref{zeta-illustration}). 
This correlation serves to magnify variance as the spatial scale increases, thus representing resources that are of intermediate clustering (e.g. fruit or small prey) to those that are highly clustered (large prey). 
The effect of extreme values of $\zeta=1$ and $\zeta=2$ on the coefficient of variation (CV) of resources across different spatial scales is shown in Fig.~\ref{zeta-illustration}.
While the meaning of $\zeta$ is layered, we show that it can be estimated empirically by the spatial distribution of resources (see Materials \& Methods and SI Appendix B).
We also show that $\zeta$ can be connected to previous explorations of home-range scaling and the fractal dimension of resource landscapes.
Preliminary analyses suggest that $\zeta$ appears to be distinct across different consumer groups, and constant across body sizes within consumer groups (SI Appendix C).

\begin{figure}
\includegraphics[width=0.4\textwidth]{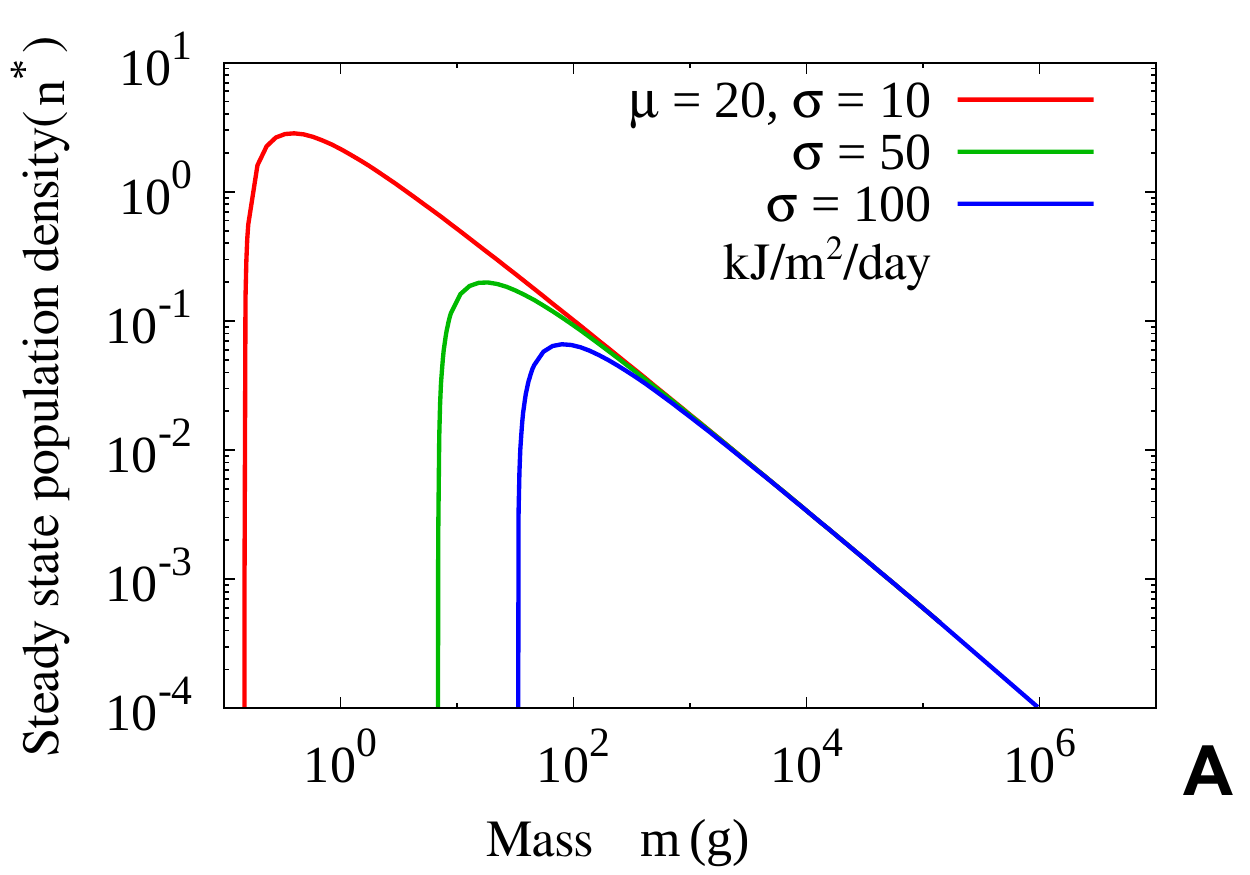}
\includegraphics[width=0.4\textwidth]{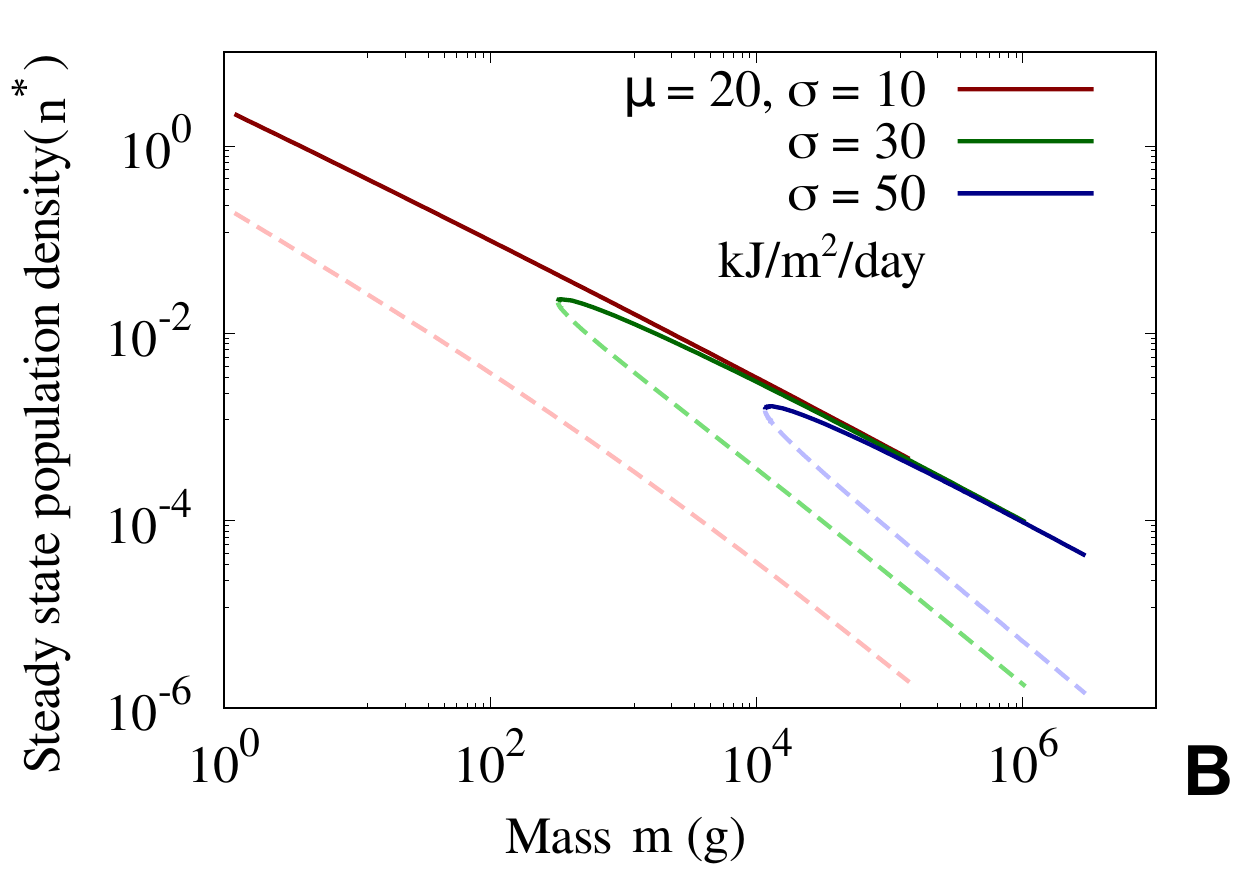}
\caption{Steady-state population densities as a function of species size for a fixed resource density ($\mu$), across various values of standard deviation ($\sigma$), (Top) for $\zeta=1$ and (Bottom) for $\zeta=2$. The dashed lines in the bottom panel show the unstable critical point $n^\circ$}
\label{steady-state-population-densities}
\end{figure}


Incorporating the above assumptions, we obtain an implicit formula to calculate the steady state population density $n^*$, given by
\begin{align}
W\left(\frac{\mu }{n^*},\frac{\sigma^2}{{n^*}^\zeta} \Bigg| r,s,\ell,b \right) = 1.
\label{population-dynamics}
\end{align}
The availability and variability of resources directly influence the profitability of different life history strategies, and ultimately determines the population density required for replacement (Eq. \ref{population-dynamics}). 
In order to gain insight into the influence of resource distributions on the expected population steady states for mammals, we assume known allometric scaling relationships for the life history parameters $r$, $s$, and $\ell$, as well as metabolic cost $b$ \cite{Brown_Towards_a_Metabolic_Theory_of_Ecology,Lindstedt_Allometry_Physiology_of_mammals,Blueweiss_Relationships_between_body_size_and_life_history}. 
We then numerically solve Eq. \ref{population-dynamics} to obtain the steady state population densities $n^*$ as a function of mass $m$ and the resource parameters $\mu$, $\sigma$ and $\zeta$.

\begin{table}[t]
\caption{Parameter definitions used for the general model and allometric scaling relationships used for parameterizing mammalian populations}
\centering
\begin{tabular}{l l l l}
\hline
Definition & Parameter & Unit\\
\hline
Variance scaling & $\zeta$ &  & \\
Energetic gain mean & $\mu$ & $\text{kJ}~\text{m}^{-2}~\text{day}^{-1}$ & \cite{Scurlock_Estimating_net_primary_productivity_from_grassland} \\
Energetic gain variance & $\sigma^2$ & $\text{kJ}^2~\text{m}^{-2\zeta} \text{day}^{-1}$ & \\ 
Metabolic scaling & $b = 5 m^{3/4}$ & $\text{kJ}$ & \cite{Brown_Towards_a_Metabolic_Theory_of_Ecology}\\
Fat energy stores scaling & $s = 0.6 m^{1.19}$ & $\text{kJ}$ & \cite{Lindstedt_Allometry_Physiology_of_mammals}\\
Reproductive scaling & $r=20 m^{0.82}$ & $\text{kJ}$ & \cite{Blueweiss_Relationships_between_body_size_and_life_history}\\
Litter size scaling & $\ell=5.71 m^{-0.10}$ &  & \cite{Blueweiss_Relationships_between_body_size_and_life_history}\\
\hline
\end{tabular}
\label{table:parameters}
\end{table}


Consumers of different body sizes feeding from different parts of the food web operate under the constraints of different resource distributions.
It is the frequency and variability with which they encounter their foods that -- in part -- structures the landscape of risk they experience.
From Eq. \ref{population-dynamics}, we can now directly assess the fitness trade-offs associated with alternative consumer foraging strategies, from grazing on homogenously distributed foods, to browsing or predating on clustered foods.

We observe that there is one stable population steady state $n^*$ for $\zeta=1$, and both a stable and an unstable steady state for $\zeta>1$ ($n^*$ and $n^\circ$, respectively).
When the resource standard deviation $\sigma$ is much smaller than the mean $\mu$, the steady state population densities follow the $-3/4$-power scaling predicted by Damuth's law \cite{Damuth1987}, given that $n^* = \mu / b_0 m^{3/4}$ for all values of $\zeta$ (Fig.~\ref{steady-state-population-densities}A).
As variance increases, $n^*$ is diminished, and this reduction disproportionately affects consumers of smaller body size (Fig.~\ref{steady-state-population-densities}A). 
When resource acquisition is more variable, the populations of larger-bodied consumers better absorb the negative effects of variable foraging success, whereas the populations of smaller-bodied consumers cannot.
This finding agrees with expectations from the fasting endurance hypothesis \cite{Lindstedt1985} and recent perspectives on the fitness benefits associated with the evolution of larger body size \cite{Yeakel_Dynamics_of_Starvation_and_Recovery}. 

\begin{figure}[t]
  \centering
  \includegraphics[width=0.4\textwidth]{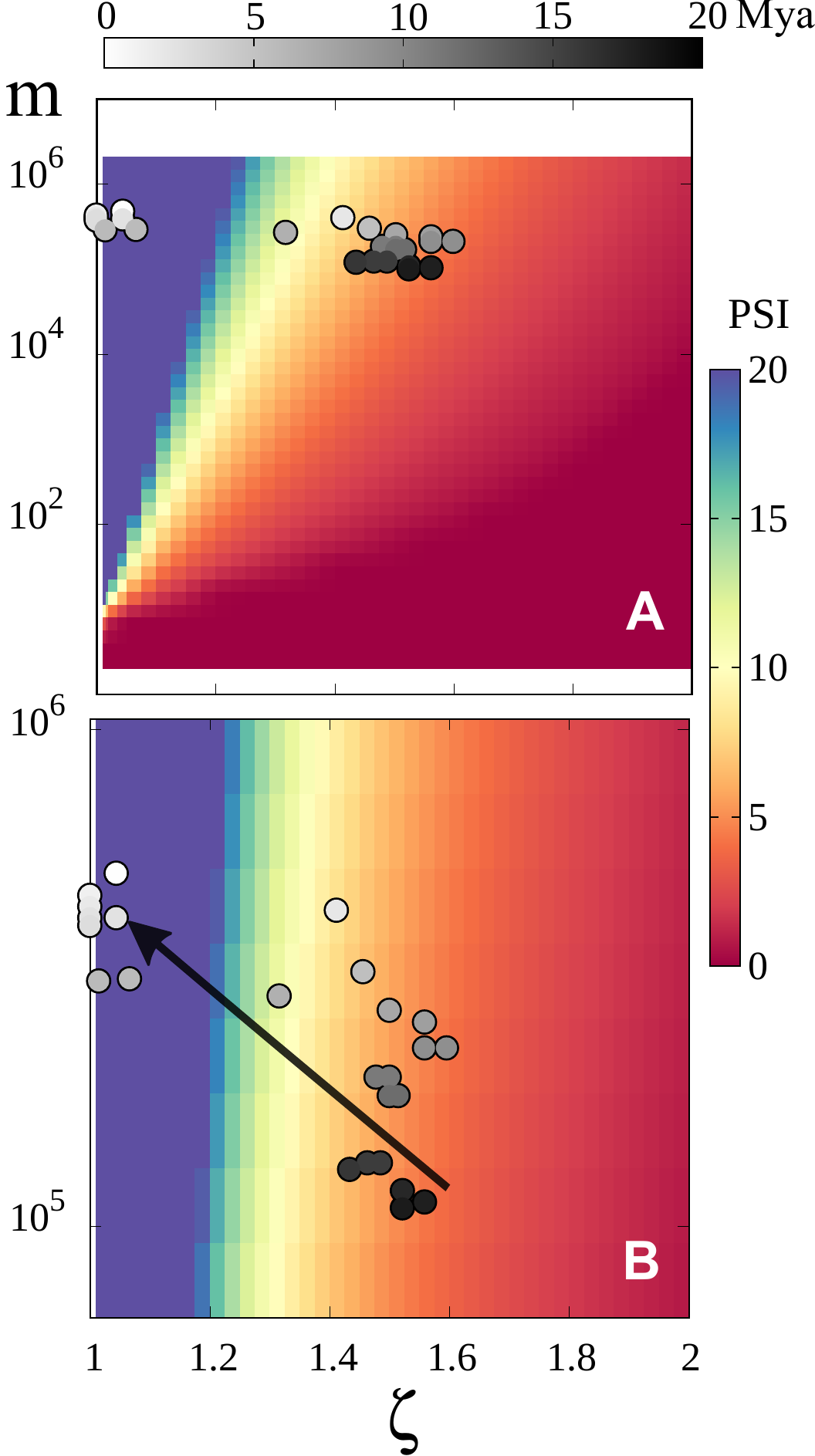}
  \caption{Population stability index (ratio of critical population density to stable population density) as a function of mass of the consumer and resource clustering, $\zeta$ for representative values of $\mu = 20 ~\text{kJ}~\text{m}^{-2}~\text{day}^{-1}$ and $\sigma^2 = 2500 ~\text{kJ}^2~\text{m}^{-2\zeta} \text{day}^{-1}$ \cite{Scurlock_Estimating_net_primary_productivity_from_grassland}. Hypothetical evolution trajectories to escape population instability are shown by the black arrow. Horse size and diet evolution is shown by the grey dots \cite{Macfadden_1996_Mammal_ancient_feeding_ecology}.} \label{fig:PSI}
\end{figure}

When resources are uniform ($\zeta=1$), a consumer population decline will ensure that more resources are available to the existing members of the population, increasing the growth rate and allowing the population to recover to the stable state. 
When resources are clustered ($\zeta>1$) we observe the appearance of a strong Allee effect where there exists a critical consumer population density $n^\circ$ below which growth rates are negative and population collapse is inevitable (Fig.~\ref{steady-state-population-densities}B). 
The solution where the stable and unstable fixed points intersect across $m$, given by $n^*(m_{\rm min})=n^\circ(m_{\rm min})$, defines the smallest consumer mass, $m_{\rm min}$, that can sustain a population in the resource environment provided by $\mu$ and $\sigma$. 

Consumers specializing on clustered resources ($\zeta>1$), will have more resources available to each individual on average (because $n^*$ is lower), but fluctuations in the population will also be larger. 
Unlike in the case of $\zeta=1$, when $\zeta>1$ both positive and negative fluctuations can decrease the growth rate to a value below unity. When the growth rate of the population falls below one and the population density is below $n^*$, the population moves below the unstable steady state $n^\circ$, resulting in collapse.

These findings reveal that increased variability in resource acquisition is expected to disproportionately affect smaller-bodied consumers that feed on clustered resources.
This suggests that if a consumer is smaller-bodied, specialization on foods that are spatially clustered carries with it larger demographic risks, whereas targeting more evenly distributed resources may provide certain selective advantages.
For small mammals such as rodents, these risks may be expected to promote the evolution of behaviors such as caching.
Because most resources, including seeds, tend to be clustered at smaller spatial scales \cite{Price1987}, caching effectively reduces the temporal variability of resource acquisition \cite{VanderWall1990}.
Consumption of spatially clustered foods that cannot be cached is generally carried out by larger consumers: for example, terrestrial frugivores tend to be larger-bodied \cite{Milton1976}, unless they are also capable of flight.
Because flight effectively increases a consumer's home-range \cite{Calder1996}, our framework predicts an increased tolerance among such consumers for spatially clustered foods, such as those fruits consumed by avian frugivores \cite{Janson1983,Herrera1989}.
Similarly, small carnivores generally specialize on smaller, more evenly distributed prey, whereas larger carnivores concentrate effort on larger, spatially clustered prey \cite{Sinclair2003}.

%
%
\section*{The adaptive benefits of grazing}
\label{sec:mammals}
The viability of a population requires that the stable steady state is sufficiently far from the unstable steady state, or critical population density, such that fluctuations will not result in collapse. 
The ratio of the stable population density to the critical population density (population stability index, ${\rm PSI}=n^*/n^\circ$) defines the stability of the population, where large values indicate that the steady state is far from the critical state, promoting stability (Fig. \ref{fig:PSI}). 
The changes in PSI as a function of consumer mass and resource clustering can be interpreted as a fitness surface for consumers of different sizes foraging on resources with different spatial distributions (color gradient in Fig. \ref{fig:PSI}).

Overall, we find that consumer populations foraging on homogeneous resources (low $\zeta$) have a higher PSI than those foraging on clustered resources (high $\zeta$), and that populations of larger-bodied consumers have greater tolerance for resources that are increasingly clustered.
If we examine this fitness surface across each dimension ($m$ and $\zeta$), we observe two directional effects:
\emph{i}) for a consumer foraging on a resource with a given spatial clustering, an increase in body mass increases PSI;
\emph{ii}) for a consumer of a given body mass, foraging on a resource with lower spatial clustering increases PSI.
Combined, these effects point to a selection gradient favoring larger-bodied consumers specializing on resources that are more homogeneously distributed.


The results of our population-level analysis suggest that differences in population stability could contribute to the selective pressures that shaped the evolution of grazing.
To test this hypothesis, we examine whether the evolution of Equidae in North America may have followed a trajectory in line with the adaptive advantages predicted by PSI.
For horse ancestors spanning a range in body mass from the early Miocene (ca. 20 Myrs BP) to the present, we estimated the scaling of horse resource clustering based on estimated reliance on graze vs. browse from the $\rm {}^{13}C/{}^{12}C$ carbon isotope ratios of fossil horse teeth \cite{Macfadden_1996_Mammal_ancient_feeding_ecology}.
We assumed that a diet of 100\% graze would result in a resource $\zeta \approx 1$, whereas a diet of 100\% browse would result in a resource $\zeta=1.72$, which we calculated from distributions of browse resources in contemporary savanna-woodland environments (see Materials and Methods and SI Appendix B).

We observe that changes in Equid body size and the spatial clustering of dietary resources follows a general trend towards increased population stability (increased PSI) over evolutionary time.
The trajectory from smaller-bodied browsers to larger-bodied grazers follows closely the fitness gradient estimated from PSI (Fig. \ref{fig:PSI}).
This suggests that reducing population instabilities induced by the joint effects of body size and uncertainties in acquiring adequate resources may have served as a catalyst for the evolutionary trend observed among Equidae in North America.
The correlation between the evolution of larger body size and increased grazing during the Miocene and Pliocene is well-known \cite{Potts1992,Bobe2006} and has been documented for many herbivore lineages on multiple continents \cite{Cerling2003,Cerling2015,Uno2011}.
Once homogeneously distributed resources such as those in grasslands became a dominant feature in Miocene environments \cite{Stromberg2011}, mammalian lineages evolved to capitalize on this relatively new food. 
While this classic evolutionary transition has been well-documented, the forces promoting increases in body size and grazing are less well understood.
We propose that the vulnerabilities associated with body size and the clustering of resource distributions provides an underlying ecological mechanism for the observed trends in the fossil record. 
Moreover, the positive relationship between PSI and body size points to a potential selective mechanism behind the evolutionary trend towards larger body size known as Cope's rule \cite{Stanley_1973_Copes_rule,Alroy1998,Clauset_2009_Body_Mass_Diversification,Smith2011}.
This finding compliments recent theoretical arguments showing that the dynamics of starvation and recovery  offer a plausible mechanistic driver of Cope’s rule \cite{Yeakel_Dynamics_of_Starvation_and_Recovery}.

Perhaps compellingly, the fitness surface that we derive by calculating PSI also reveals that there should be increased tolerance for specialization on clustered resources among larger-bodied consumers.
While the attainment of the large-bodied grazing niche evolved quickly within the relatively smaller Equidae, the change was much slower for the larger-bodied Gomphotheres and Elephantids, and did not occur at all for Deinotheres and Giraffids \cite{Uno2011}.
Many other selective pressures that we do not address -- such as interspecific competition and predation -- undoubtedly had an enormous influence on the evolutionary trajectories of grazers.
We suggest that the fitness differences arising from the foraging risks arising from the exploitation of clustered resources may have contributed to observed macroevolutionary trends among mammals following the advent of grasslands.

The risk landscape experienced by consumers specializing on different resource types largely depends on body size.
However body size also influences the risk landscape by determining the area over which resources are foraged, and consequently, the spatial clustering of those resources experienced by the consumer.
As we have shown, the availability and variability of resources is expected to have a large influence on optimal life history strategies and population-level stability.
Explicitly incorporating variability into models of resource acquisition reveals ecological constraints that may have played an important role in observed evolutionary trends associated with correlated changes in consumer body size and resource type.
In the future, accounting for the dynamics of individual state, combined with introducing more complex foraging behaviors, may provide additional insight into the ecological constraints influencing the evolutionary trajectories of species.

\section*{Materials \& Methods}
\begin{footnotesize}
  \subsection*{Estimating $\zeta$ from resource landscape satellite image}
We estimate the value of $\zeta$ for a clustered resource such as browse by sampling the spatial distributions of trees and shrubs from satellite images of the serengeti. Specifically, high-resolution images from coordinates $-4^\circ 55' 20.93'',37^\circ 0' 30.32''$ are retrieved from \textit{Google Earth} (see S.I.). Image sizes are of either $1000$x$1000$ or $2000$x$2000$ pixels. These images are first converted to grayscale and the pixel intensities are transformed to binary values using a threshold value selected to best match the tree/shrub distribution observed visually, thus producing a matrix of presence or absence of resource. The value of $\zeta$ is then estimated as follows. For each integer $x \in [10,100]$ such that $1 \ll x \ll 1000$, a sample ($N=10^4$) of boxes are chosen. The variance of total tree-pixels within each box is calculated across this sample, and for each value of $x$. The slope of $\log(\text{variance})$ vs. $\log(x^2)$ provides the value of $\zeta$. The average (standard deviation) of $\zeta$ across four satellite images is $1.72~(0.07)$. The value of $\zeta$ for a uniform patch of grass (modeled as a uniform density across the given area) is $\zeta=1$ from the central limit theorem \cite{Durrett_2019_Probability}. We use the two extreme values ($\zeta=1$ and $\zeta=1.72$) as the values of $\zeta$ for all grazing and all browsing behaviors respectively. Paleontological dietary compositions for horses is estimated by using $\delta^{13}C$ isotope measurements from \cite{Macfadden_1996_Mammal_ancient_feeding_ecology}. All C3 diet is assumed to have $\zeta=1.72$ equal value for the distribution of trees and shrubs we measure on the serengeti. All C4 diet is assumed to have $\zeta=1$ equal to the expected value for a uniform distribution of grass. Mixed diets are assumed to have a linearly interpolated value of $\zeta$ between $1$ and $1.72$.
\end{footnotesize}


\clearpage

\beginsupplement

\section*{Supplementary Information}
\subsection*{Appendix A}
Here we derive formula given in Eq.~1 of the main text, reproduced below,
\begin{equation}
 \left\langle W \right\rangle = \frac{\ell\left(1-\exp\left(-2(\mu-b)\epsilon r/\ell \sigma^2\right)\right)}{\exp\left(-2(\mu-b)s/\sigma^2\right)-\exp\left(-2(\mu-b)(s+r)/\sigma^2\right)}\,.
 \label{solution}
\end{equation}
To derive Eq.~\ref{solution}, we first define the following probabilities,
\begin{enumerate}
\item $\text{P}_0 = $ Probability that a diffusive trajectory with bias $\mu$ and variance $\sigma^2$ reaches reproductive threshold $x=s+r$, when starting from the initial state $x = \epsilon r / \ell$,
\item $\text{Q}_0 = 1 - \text{P}_0 = $ Probability that a diffusive trajectory reaches starvation threshold $x=0$ when starting from the initial state $x = \epsilon r /\ell$,
\item $\text{P} = $ Probability that a diffusive trajectory reaching reproductive threshold after a reproductive event (i.e., starting from a state $x = s$),
\item $\text{Q} = $ Probability of starving after a reproductive event.
\end{enumerate}
\begin{figure}[hb]
\centering
\includegraphics[width=0.5\textwidth]{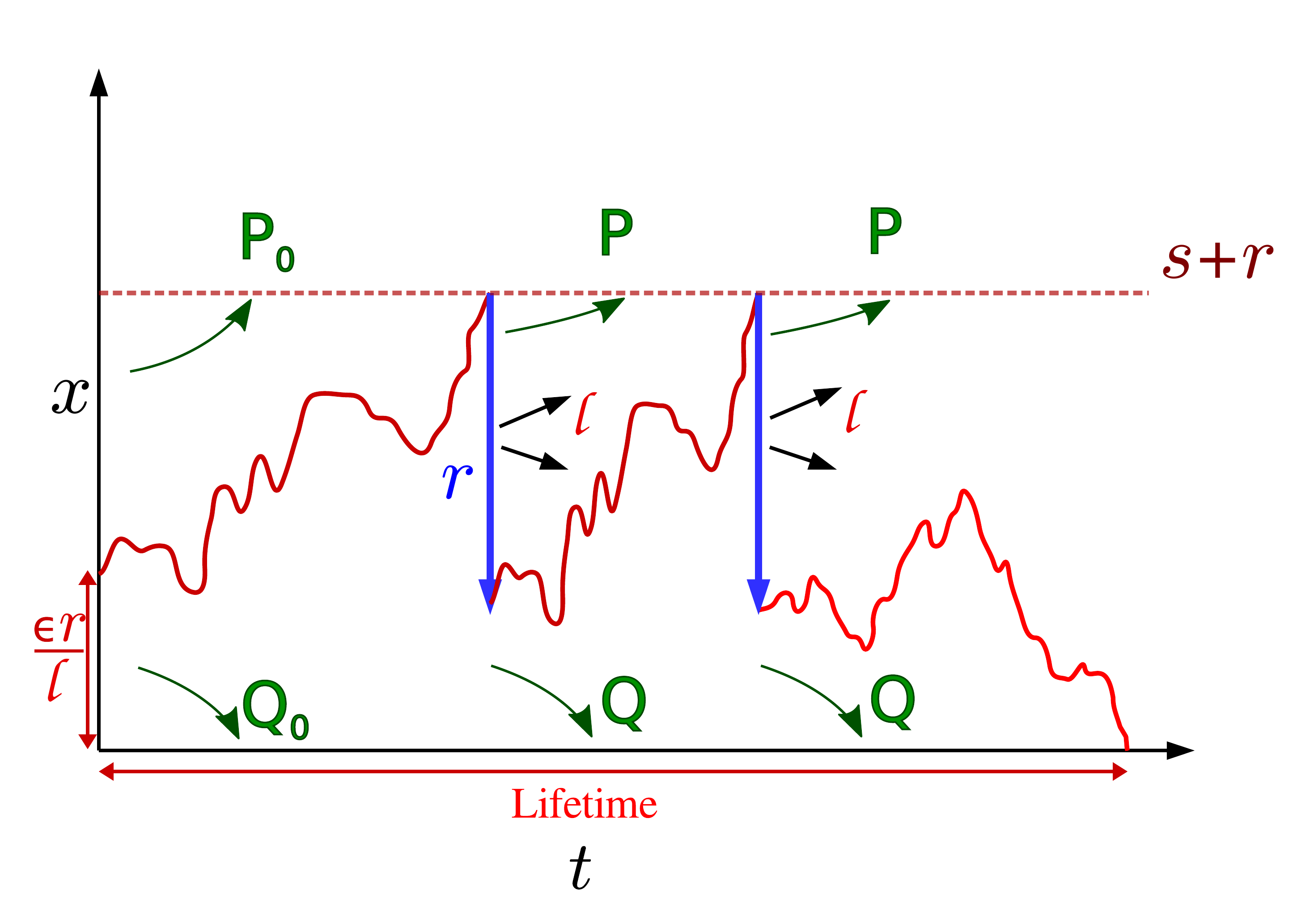}
\caption{A cartoon showing the relevant probabilities $\text{P}_0$, $\text{Q}_{0}$, P and Q.}
\end{figure}
With these probabilities, we can write the average number of offsprings and its standard deviation as follows,
\begin{align}
\left\langle W \right\rangle &= 0\cdot \text{Q}_0 + \ell \cdot \text{P}_0 \text{Q} + 2\ell\cdot \text{P}_0 \text{P}\text{Q} + 3\ell\cdot \text{P}_0 \text{P}^2\text{Q} + \hdots \\ \nonumber
&= \ell~\text{P}_0 / \text{Q}
\end{align}
\begin{align}
\left\langle W^2 \right\rangle &= 0^2 \cdot \text{Q}_0 + \ell^2\cdot \text{P}_0 \text{Q} + (2\ell)^2\cdot \text{P}_0 \text{P}\text{Q} \\ \nonumber
&+ (3\ell)^2\cdot \text{P}_0 \text{P}^2\text{Q} + \hdots \\ \nonumber
&= \ell^2~\text{P}_0(1+\text{P}) / \text{Q}^2
\end{align}
Using the first two moments, the variance is given by,
\begin{equation}
\mathrm{Var}(W) = \left\langle W^2 \right\rangle-\left\langle W \right\rangle^2 = \ell^2\,\text{P}_0(1+\text{P}-\text{P}_0) / \text{Q}^2
\label{wsimple}
\end{equation}
\begin{equation}
\mathrm{Stdev}(W) = \sqrt{\mathrm{Var}(r)} = \ell\sqrt{\text{P}_0(1+\text{P}-\text{P}_0)} / \text{Q}
\label{stdevwsimple}
\end{equation}
Where,
\begin{equation}
\text{P}_0 = \frac{1-\exp{\left(-2(\mu-b)\epsilon r /\ell \sigma^2\right)}}{1-\exp{\left(-2(\mu-b)(s+r)/\sigma^2\right)}}
\label{p0}
\end{equation}
\begin{equation}
\text{Q}_0 = 1-\text{P}_0 = \frac{\exp{\left(-2(\mu-b)\epsilon r /\ell\sigma^2\right)}-\exp{\left(-2(\mu-b)(s+r) / \sigma^2\right)}}{1-\exp{\left(-2(\mu-b)(s+r)/\sigma^2\right)}}
\label{q0}
\end{equation}
\begin{equation}
\text{P} = \frac{1-\exp{\left(-2(\mu-b)s /\sigma^2\right)}}{1-\exp{\left(-2(\mu-b)s/\sigma^2\right)}}
\label{p1}
\end{equation}
\begin{equation}
\text{Q} = 1-\text{P} = \frac{\exp{\left(-2(\mu-b)s /\sigma^2\right)}-\exp{\left(-2(\mu-b)(s+r) / \sigma^2\right)}}{1-\exp{\left(-2(\mu-b)(s+r)/\sigma^2\right)}}
\label{q1}
\end{equation}
~\eqref{p0}--~\eqref{q1} are obtained from \cite{Redner_2001_Guide_FPP}. Substituting the above expressions in ~\eqref{wsimple} and \eqref{stdevwsimple}, we get,
\begin{equation}
 \left\langle W \right\rangle = \frac{\ell\left(1-\exp\left(-2(\mu-b)\epsilon r/\ell \sigma^2\right)\right)}{\exp\left(-2(\mu-b)s/\sigma^2\right)-\exp\left(-2(\mu-b)(s+r)/\sigma^2\right)}\,.
\end{equation}
and,
\begin{equation}
 \mathrm{Stdev}(W) = \frac{\ell\sqrt{
 \begin{aligned}
 &\left(1-\exp\left(-2(\mu-b)\epsilon r/\ell \sigma^2\right)\right)\\
 &\times \left(1-\exp{\left(-2(\mu-b)(s+r)/\sigma^2\right)}\right. \\&\left.-\exp{\left(-2(\mu-b)s / \sigma^2\right)}\right.\\
 &\left.+\exp{\left(-2(\mu-b)\epsilon r / \sigma^2\right)}\right)
 \end{aligned}
 }}{\exp\left(-2(\mu-b)s/\sigma^2\right)-\exp\left(-2(\mu-b)(s+r)/\sigma^2\right)}\,.
\end{equation}

\subsection*{Appendix B}
\begin{figure*}[h]
\includegraphics[width=1\textwidth]{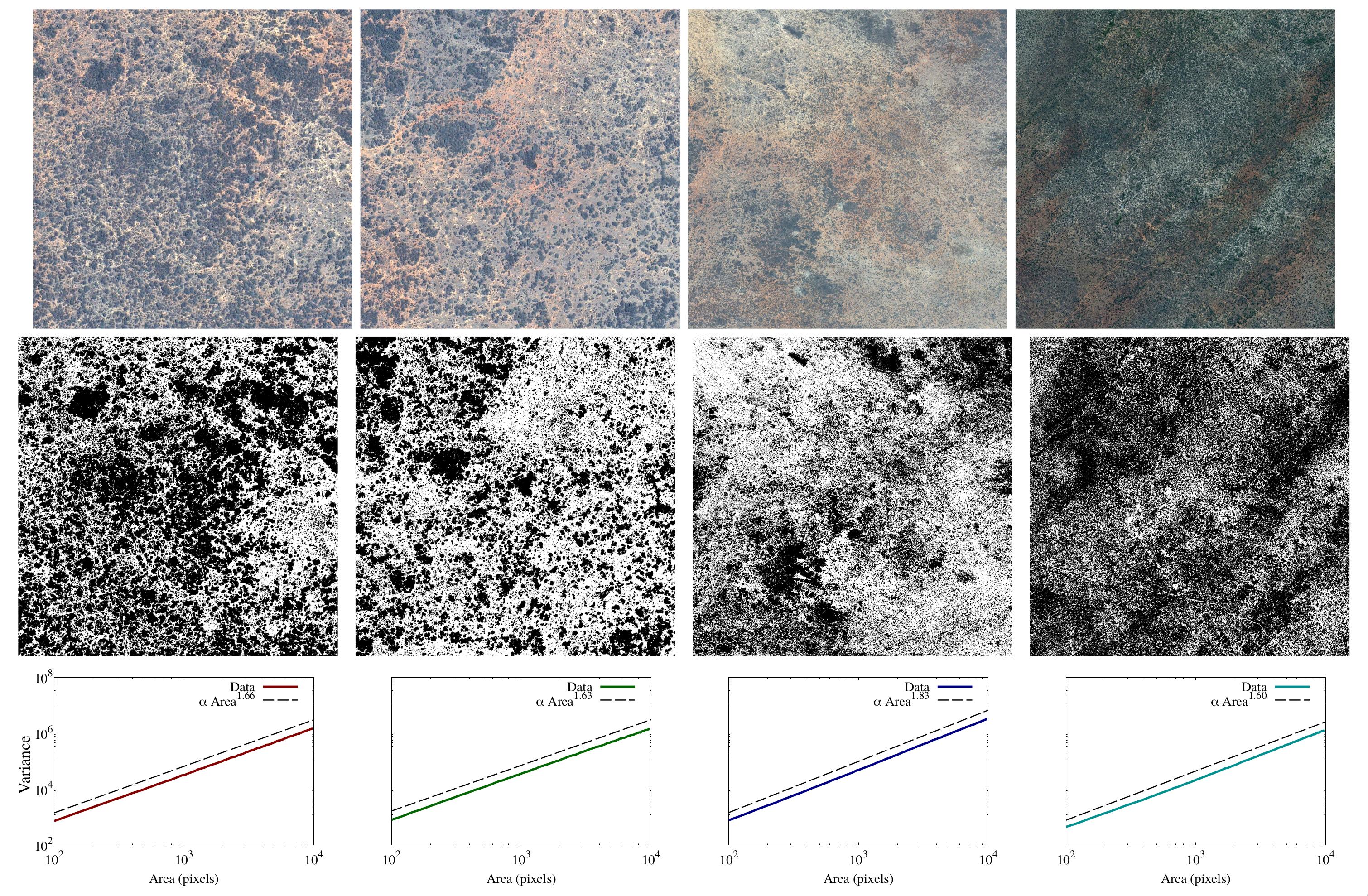}
\caption{(Top) satellite images of a Savannah landscape. (Middle) The images after gray-scaling and thresholding the pixel intensities reveal tree-locationn as black pixels and grass/dirt as white pixels. The value of threshold is $(150/256)$ for first three images and $100/256$ for the rightmost image (Bottom) Plots showing the scaling of variance of resources with the area. The two images to the left have a resolution of $1000$x$1000$ pixels, and the two images to the right have a resolutionn of $2000$x$2000$ pixels.}
\label{googleearth}
\end{figure*}

Much of our model centers on the variance scaling exponent $\zeta$ which parameterizes how the variance faced by an individual changes as a function of population density. The daily availability of resources to an individual consumer is assumed to be proportional to the total available biomass in the consumer's homerange. The value of $\zeta$ is estimated by measuring the variance in resources as a function of its homerange area. We obtain the biomass distribution in a landscape using a high-resolution satellite images obtained from \textit{Google Earth} (see fig.~\ref{googleearth}). The value of $\zeta$ can be measured from an image using the following steps,
\begin{enumerate}
\item Convert the image to grayscale.
\item Apply thresholding to the image to produce pixels that are either black or white. The value of threshold is chosen to match visually the tree distribution seen in the original image.
\item Produce a matrix of zeros and ones from the black-and-white image. We use the \texttt{ImageData} function from Mathematica.
\item Choose a range of box sizes, $x$, such that $1\ll x \ll L$ where $L$x$L$ pixels is the size of the image. We use images with $L=1000$ and $L=2000$. We use $x \in [10,100]$.
\item For each box size $x$, choose a sample $N$ of boxes at random from the image. We use $N = 10^4$.
\item Count the total number of resource pixels in each box, and calculate the variance across the sample.
\item Plot $\log(\text{variance})$ vs $\log(x)$ and measure its slope. This will be the value of $\zeta$ for the landscape in the image.
\end{enumerate}

We conducted the following sensitivity analyses. Changing the threshold value by $20\%$ changes the the measured value of $\zeta$ by less than $2\%$ indicating that $\zeta$ is not very sensitive to the choice of threshold. Using a subset of $1000$x$1000$ pixels cropped from the third and fourth images reduces the measured $\zeta$ by roughly $10\%$ indicating the presence of finite-size effects.

\subsection*{Appendix C}
 There have been a variety of previous efforts centered on understanding the spatial requirements for an individual to obtain enough resources on average. These efforts have focused on the fractal dimension of landscapes and the spatial distribution of individuals along with the scaling of home range area as function of body size. Previous predictions for the scaling of home range size with body size have invoked arguments related to fractal dimension \cite{haskell2002fractal} or the rate of interaction amongst members of the same species \cite{jetz2004scaling}. Both sets of arguments are fundamentally about the requirements for resource gathering in order to meet metabolic needs on average, but make different assumptions about what adjusts resource availability on a landscape. 

%
%

The previous work \cite{haskell2002fractal} related to the average resource intake of a forager on a fractal landscape has argued that home range size should scale according to
\begin{equation}
H\propto M^{\alpha +\eta\left(D-F\right)} 
\end{equation}
where $\alpha$ is the metabolic scaling exponent, $\eta$ is the scaling exponent for a typical length scale with body size (e.g. stride length), $D$ is the dimension of the environment occupied by a class of organisms (e.g. $2$ for terrestrial mammals and $3$ for aquatic organisms),and $F$ is the fractal dimension of the environment. If we define the scaling exponent between home range and body mass to be $b$ then we have that 
\begin{equation}
F=\frac{\alpha}{\eta}+D-\frac{b}{\eta}
\end{equation}
and thus for fixed values of $\alpha$, $D$, and $\eta$ the home range scaling is consistent with a single fractal dimension.  Empirically, the scaling between home range and body mass has been found to be $b=0.83$ for terrestrial mammalian herbivores and $b=1.21$ for terrestrial mammalian carnivores. Taking $D=2$, $\alpha=3/4$, and $\eta=1/3$ (for the scaling of stride lengths in terrestrial mammals), then we have that $F=17/4-3b$ and that $F=1.76$ for mammalian herbivores and $F=0.62$ for mammalian carnivores based on the measured home range scaling. These results are consistent with the idea that mammals of all body sizes experience a landscape defined by a common fractal dimension, and suggests that a single $\zeta$ may be appropriate for describing mammals of different body size but similar trophic strategy.


\begin{thebibliography}{10}

\bibitem{Mangel1986}
Mangel M, Clark CW (1986) {Towards a unified foraging theory}.
\newblock {\em Ecology} 67(5):1127--1138.

\bibitem{Roff2002}
Roff D (2002) {\em Life History Evolution}.
\newblock (Sinauer).

\bibitem{Gerber2004}
Gerber LR, Reichman OJ, Roughgarden J (2004) {Food hoarding: future value in
  optimal foraging decisions}.
\newblock {\em Ecol. Model.} 175(1):77--85.

\bibitem{Yeakel_Dynamics_of_Starvation_and_Recovery}
Yeakel JD, Kempes CP, Redner S (2016) The dynamics of starvation and recovery.
\newblock {\em arXiv preprint arXiv:1608.08995}.

\bibitem{Lindstedt1981}
Lindstedt SL, W~A~Calder sIs (1981) {Body Size, Physiological Time, and
  Longevity of Homeothermic Animals}.
\newblock {\em Q Rev Biol} 56(1):1--16.

\bibitem{Burke_Central_place_foraging_response_to_fluctuations}
M. BC, A. MW (2009) The foraging decisions of a central place foraging seabird
  in response to fluctuations in local prey conditions.
\newblock {\em Journal of Zoology} 278(4):354--361.

\bibitem{Schmickl_Cost_of_Fluctuations_in_Honey_Bees}
Schmickl T, Crailsheim K (2004) Costs of environmental fluctuations and
  benefits of dynamic decentralized foraging decisions in honey bees.
\newblock {\em Adaptive Behavior} 12(3-4):263--277.

\bibitem{Brown_Marquet_Evolution_of_Body_Size}
Brown JH, Marquet PA, Taper ML (1993) Evolution of body size: Consequences of
  an energetic definition of fitness.
\newblock {\em The American Naturalist} 142(4):573--584.
\newblock PMID: 19425961.

\bibitem{Grunbaum2012}
Gr{\"u}nbaum D (2012) {The logic of ecological patchiness.}
\newblock {\em Interface Focus} 2(2):150--155.

\bibitem{Charnov_Optimal_foraging}
Charnov EL (1976) Optimal foraging, the marginal value theorem.
\newblock {\em Theoretical population biology} 9(2):129--136.

\bibitem{MacArthur_Pianka_On_Optimal_Use_of_a_Patchy_Environment}
MacArthur RH, Pianka ER (1966) On optimal use of a patchy environment.
\newblock {\em The American Naturalist} 100(916):603--609.

\bibitem{vanNoordwijk1986}
van Noordwijk AJ, de~Jong G (1986) {Acquisition and Allocation of Resources:
  Their Influence on Variation in Life History Tactics}.
\newblock {\em Am. Nat.} 128(1):137--142.

\bibitem{Stearns1989}
Stearns SC (1989) {Trade-offs in life-history evolution}.
\newblock {\em Funct. Ecol.} 3(3):259.

\bibitem{Charnov_1991_Life_history_variation}
Charnov EL (1991) Evolution of life history variation among female mammals.
\newblock {\em Proceedings of the National Academy of Sciences}
  88(4):1134--1137.

\bibitem{Abrams2001}
Abrams PA (2001) {Modelling the adaptive dynamics of traits involved in inter-
  and intraspecific interactions: An assessment of three methods}.
\newblock {\em Ecol. Lett.} 4:166--175.

\bibitem{Jarvis2003}
Jarvis JUM (2003) { Foraging in the subterranean social Damaraland mole-rat,
  Cryptomys damarensis : an investigation into size-dependent geophyte
  utilization and foraging patterns }.
\newblock {\em Can. J. Zool.} 81(4):743--752.

\bibitem{Fryxell2005}
Fryxell JM, et~al. (2005) Landscape scale, heterogeneity, and the viability of
  serengeti grazers.
\newblock {\em Ecology Letters} 8(3):328--335.

\bibitem{Malmborg1988}
Malmborg PK, Willson MF (1988) {Foraging Ecology of Avian Frugivores and Some
  Consequences for Seed Dispersal in an Illinois Woodlot}.
\newblock {\em The Condor: Ornithological Applications} 90(1):173--186.

\bibitem{Hatton2015}
Hatton IA, et~al. (2015) {The predator-prey power law: Biomass scaling across
  terrestrial and aquatic biomes.}
\newblock {\em Science} 349(6252):aac6284--aac6284.

\bibitem{Roff2007}
Roff DA, Fairbairn DJ (2007) {The evolution of trade-offs: where are we?}
\newblock {\em Journal of Evolutionary Biology} 20(2):433--447.

\bibitem{Moore2014}
Moore JW, Yeakel JD, Peard D, Lough J, Beere M (2014) {Life-history diversity
  and its importance to population stability and persistence of a migratory
  fish: steelhead in two large North American watersheds}.
\newblock {\em J. Anim. Ecol.} 83(5):1035--1046.

\bibitem{jetz2004scaling}
Jetz W, Carbone C, Fulford J, Brown JH (2004) The scaling of animal space use.
\newblock {\em Science} 306(5694):266--268.

\bibitem{DeLong2012a}
DeLong JP, Vasseur DA (2012) {Size-density scaling in protists and the links
  between consumer{\textendash}resource interaction parameters}.
\newblock {\em J. Anim. Ecol.} 81(6):1193--1201.

\bibitem{DeLong2012b}
DeLong JP (2012) {Experimental demonstration of a
  {\textquoteleft}rate{\textendash}size{\textquoteright} trade-off governing
  body size optimization}.
\newblock {\em Evol. Ecol. Res.} 14(3):343--352.

\bibitem{haskell2002fractal}
Haskell JP, Ritchie ME, Olff H (2002) Fractal geometry predicts varying body
  size scaling relationships for mammal and bird home ranges.
\newblock {\em Nature} 418(6897):527.

\bibitem{Pawar_2012_dimensionality}
Pawar S, Dell AI, Savage VM (2012) Dimensionality of consumer search space
  drives trophic interaction strengths.
\newblock {\em Nature} 486(7404):485.

\bibitem{Redner2014}
Metzler R, Oshanin G, Redner S (2014) {\em First-passage phenomena and their
  applications}.
\newblock (World Scientific).

\bibitem{Damuth1987}
Damuth J (1987) {Interspecific allometry of population density in mammals and
  other animals: the independence of body mass and population energy-use}.
\newblock {\em Biol. J. Linn. Soc.} 31(3):193--246.

\bibitem{Mangel_Unified_Foraging_Theory}
Mangel M, Clark CW (1986) Towards a unifield foraging theory.
\newblock {\em Ecology} 67(5):1127--1138.

\bibitem{Hilborn_1997_Ecological_Detective}
Hilborn R, Mangel M (1997) {\em The ecological detective: confronting models
  with data}.
\newblock (Princeton University Press) Vol.{}~28.

\bibitem{Reznick2002}
Reznick D, Bryant MJ, Bashey F (2002) {r- and K-selection revisited: The role
  of population regulation in life-history evolution}.
\newblock {\em Ecology} 83(6):1509--1520.

\bibitem{Crespi2002}
Crespi BJ, Teo R (2002) Comparative phylogenetic analysis of the evolution of
  semelparity and life history in salmonid fishes.
\newblock {\em Evolution} 56(5):1008--1020.

\bibitem{Pianka1970}
Pianka ER (1970) On r- and k-selection.
\newblock {\em Am. Nat.} 104(940):592--597.

\bibitem{Rood_1990_Group_size_survival_reproduction}
Rood JP (1990) Group size, survival, reproduction, and routes to breeding in
  dwarf mongooses.
\newblock {\em Animal Behaviour} 39(3):566--572.

\bibitem{Cahan2002}
Cahan SH, Blumstein DT, Sundström L, Liebig J, Griffin A (2002) Social
  trajectories and the evolution of social behavior.
\newblock {\em Oikos} 96(2):206--216.

\bibitem{Taylor1961}
Taylor LR (1961) Aggregation, variance and the mean.
\newblock {\em Nature} 189(4766):732--735.

\bibitem{Brown_Towards_a_Metabolic_Theory_of_Ecology}
Brown JH, Gillooly JF, Allen AP, Savage VM, West GB (2004) Toward a metabolic
  theory of ecology.
\newblock {\em Ecology} 85(7):1771--1789.

\bibitem{Lindstedt_Allometry_Physiology_of_mammals}
Lindstedt SL, Schaeffer PJ (2002) Use of allometry in predicting anatomical and
  physiological parameters of mammals.
\newblock {\em Laboratory Animals} 36(1):1--19.
\newblock PMID: 11833526.

\bibitem{Blueweiss_Relationships_between_body_size_and_life_history}
Blueweiss L, et~al. (1978) Relationships between body size and some life
  history parameters.
\newblock {\em Oecologia} 37(2):257--272.

\bibitem{Scurlock_Estimating_net_primary_productivity_from_grassland}
Scurlock JMO, Johnson K, Olson RJ (2002) Estimating net primary productivity
  from grassland biomass dynamics measurements.
\newblock {\em Global Change Biology} 8(8):736--753.

\bibitem{Lindstedt1985}
Lindstedt SL, Boyce MS (1985) {Seasonality, Fasting Endurance, and Body Size in
  Mammals}.
\newblock {\em Am. Nat.} 125(6):873--878.

\bibitem{Macfadden_1996_Mammal_ancient_feeding_ecology}
Macfadden BJ, Cerling TE (1996) Mammalian herbivore communities, ancient
  feeding ecology, and carbon isotopes: A 10 million-year sequence from the
  neogene of florida.
\newblock {\em Journal of Vertebrate Paleontology} 16(1):103--115.

\bibitem{Price1987}
Price MV, Reichman OJ (1987) Distribution of seeds in sonoran desert soils:
  Implications for heteromyid rodent foraging.
\newblock {\em Ecology} 68(6):1797--1811.

\bibitem{VanderWall1990}
Vander~Wall S (1990) {\em Food Hoarding in Animals}.
\newblock (University of Chicago Press).

\bibitem{Milton1976}
Milton K, May ML (1976) Body weight, diet and home range area in primates.
\newblock {\em Nature} 259(5543):459--462.

\bibitem{Calder1996}
Calder W (1996) {\em Size, Function, and Life History}, Doover books on
  biology, psychology and medicine.
\newblock (Dover Publications).

\bibitem{Janson1983}
Janson CH (1983) Adaptation of fruit morphology to dispersal agents in a
  neotropical forest.
\newblock {\em Science} 219(4581):187--189.

\bibitem{Herrera1989}
Herrera CM (1989) Vertebrate frugivores and their interaction with invertebrate
  fruit predators: Supporting evidence from a costa rican dry forest.
\newblock {\em Oikos} 54(2):185--188.

\bibitem{Sinclair2003}
Sinclair ARE, Mduma S, Brashares JS (2003) {Patterns of predation in a diverse
  predator{\textendash}prey system}.
\newblock {\em Nature} 425:288--290.

\bibitem{Potts1992}
Potts R, Behrensmeyer A (1992) Late cenozoic terrestrial ecosystems in {\em
  Terrestrial Ecosystems Through Time: Evolutionary Paleoecology of Terrestrial
  Plants and Animals}, eds.{} Behrensmeyer A, et~al.
\newblock (University of Chicago Press, Chicago), pp. 419--541.

\bibitem{Bobe2006}
Bobe R (2006) {The evolution of arid ecosystems in eastern Africa}.
\newblock {\em J. Arid Environ.} 66(3):564--584.

\bibitem{Cerling2003}
Cerling TE, Harris JM, Passey BH (2007) {Diets of east african bovidae based on
  stable isotope analysis}.
\newblock {\em Journal of Mammology} p.~15.

\bibitem{Cerling2015}
Cerling TE, et~al. (2015) {Dietary changes of large herbivores in the Turkana
  Basin, Kenya from 4 to 1 Ma.}
\newblock {\em Proc. Natl. Acad. Sci. USA} 112(37):11467--11472.

\bibitem{Uno2011}
Uno K, Cerling T, al e (2011) {Late Miocene to Pliocene carbon isotope record
  of differential diet change among East African herbivores}.
\newblock {\em Proc. Natl. Acad. Sci. USA}.

\bibitem{Stromberg2011}
Str{\"o}mberg CAE (2011) {Evolution of Grasses and Grassland Ecosystems}.
\newblock {\em Annu. Rev. Earth. Pl. Sc.} 39(1):517--544.

\bibitem{Stanley_1973_Copes_rule}
Stanley SM (1973) An explanation for cope's rule.
\newblock {\em Evolution} 27(1):1--26.

\bibitem{Alroy1998}
Alroy J (1998) {Cope's rule and the dynamics of body mass evolution in North
  American fossil mammals}.
\newblock {\em Science} 280(5364):731--734.

\bibitem{Clauset_2009_Body_Mass_Diversification}
Clauset A, Redner S (2009) Evolutionary model of species body mass
  diversification.
\newblock {\em Phys. Rev. Lett.} 102(3):038103.

\bibitem{Smith2011}
Smith FA, Lyons SK (2011) {How big should a mammal be? A macroecological look
  at mammalian body size over space and time}.
\newblock {\em Philos. T. Roy. Soc. B} 366(1576):2364--2378.

\bibitem{Durrett_2019_Probability}
Durrett R (2019) {\em Probability: theory and examples}.
\newblock (Cambridge university press) Vol.{}~49.

\end{thebibliography}

\begin{thebibliography}{1}

\bibitem{Redner_2001_Guide_FPP}
Redner S (2001) {\em A guide to first-passage processes}.
\newblock (Cambridge University Press).

\bibitem{haskell2002fractal}
Haskell JP, Ritchie ME, Olff H (2002) Fractal geometry predicts varying body
  size scaling relationships for mammal and bird home ranges.
\newblock {\em Nature} 418(6897):527.

\bibitem{jetz2004scaling}
Jetz W, Carbone C, Fulford J, Brown JH (2004) The scaling of animal space use.
\newblock {\em Science} 306(5694):266--268.

\end{thebibliography}

\end{document}